\documentclass[twocolumn,reprint,letterpaper,preprintnumbers,showpacs,amsmath,amssymb,10pt]{revtex4}

\usepackage{graphicx}% Include figure files
\usepackage{dcolumn}% Align table columns on decimal point
\usepackage{bm}% bold math
\usepackage{psfrag}
\usepackage{color}

\begin{document}

%\preprint{{\it In consideration in PRL} }

\title{Locomotion of Electrocatalytic Nanomotors due to Reaction Induced Charge Auto-Electrophoresis}% Force line breaks with \\

\author{J.L. Moran, P.M. Wheat}
\author{J.D. Posner}%
% \email{jposner@asu.edu}
\affiliation{Mechanical Engineering, Chemical Engineering, Arizona State University, Tempe, AZ 85287, USA}

\date{\today}% It is always \today, today,

\begin{abstract}
Bimetallic rod-shaped nanomotors swim autonomously in hydrogen peroxide solutions.   Here we present a scaling analysis, computational simulations, and experimental data that show that the nanomotor locomotion is driven by fluid slip around the nanomotor surface due to electrical body forces.  The body forces are generated by a coupling of  charge density and electric fields induced by electrochemical reactions occurring on the nanomotor surface.  We describe the dependence of nanomotor motion on the nanomotor surface potential and reaction-driven flux.
\end{abstract}
\maketitle

Autonomously propelled synthetic nanomotors represent a major step in the development of practical nanomachines because they are  able to perform nanoscale tasks without the need for externally supplied energy.  
%Like their naturally occurring counterparts ~\cite{waterbury_cyanobacterium_1985}, these nanomotors generate propulsive forces which drive their autonomous locomotion.  
Synthetic nanomotors often take the form of Janus nanoparticles consisting of two dissimilar segments  suspended in a liquid fuel with asymmetric reactions occurring on the nanomotor surface.  Examples include the catalytic \cite{howse_self-motile_2007,gibbs_autonomously_2009,fournier-bidoz_synthetic_2005}  and electrocatalytic~\cite{burdick_synthetic_2008,paxton_motility_2005} decomposition of hydrogen peroxide and enzymatic reactions~\cite{golestanian_propulsion_2005,mano_bioelectrochemical_2005}.  The nature of the reaction and the locomotion mechanism at work depend on the fuel and particle materials.  Several particle locomotion mechanisms have been investigated including autonomous variants of diffusiophoresis~\cite{golestanian_propulsion_2005}, osmiophoresis~\cite{golestanian_propulsion_2005,cordova-figueroa_osmotic_2008}, electrophoresis~\cite{paxton_motility_2005,wang_bipolar_2006}, surface-tension gradients~\cite{paxton_catalytic_2004}, and bubble propulsion~\cite{gibbs_autonomously_2009}.  In addition, Golestanian and coworkers have provided a general mathematical framework for phoretic swimmers based on a particle's surface activity and mobility that can be applied to several of the aforementioned mechanisms~\cite{golestanian_designing_2007}.

Several groups have fabricated self-propelled bimetallic nanowire motors that swim due to the electrochemical decomposition of hydrogen peroxide (H$_2$O$_2$) fuel to oxygen and water~\cite{paxton_motility_2005,burdick_synthetic_2008,fournier-bidoz_synthetic_2005}.  These electrochemically grown nanowires (or nanorods) have been engineered to: (i) swim at one hundred body lengths per second~\cite{laocharoensuk_carbon-nanotube-induced_2008}; (ii) perform controlled motion under applied magnetic fields~\cite{burdick_synthetic_2008,kline_catalytic_2005}, chemical\cite{calvo-marzal_electrochemically-triggered_2009} and thermal \cite{balasubramanian_thermal_2009-1} modulation; (iii) sense  chemicals  through their autonomous motion~\cite{kagan_chemical}; and (iv)  pick up, haul, and release micron-scale cargo~\cite{burdick_synthetic_2008}.  Several of the aforementioned studies have contributed conceptual models and experimental data in an effort to elucidate the mechanism by which these bimetallic nanorods convert chemical energy into motion.  However, thus far there is no universally accepted theory that fully describes the locomotion physics. 

In this Letter, we present a set of governing equations, a scaling analysis, numerical simulations, and experiments that describe the physics underlying the autonomous motion of electrocatalytic bimetallic nanomotors due to a mechanism we call Reaction Induced Charge Auto-Electrophoresis (RICA).  To our knowledge, this is the first work which solves the Poisson, advection-diffusion, and Navier-Stokes equations to provide a detailed physical description of the locomotion of bimetallic nanorods.  The scaling analysis and simulations enable predictions of nanomotor velocity, direction, and total propulsive force as a function of two relevant parameters of the system.  We compare these predictions to experimental measurements and observe excellent agreement.

\begin{figure}
\includegraphics[clip,width=2.8 in]{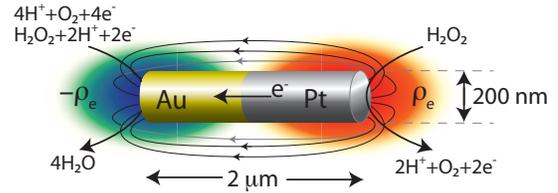} % Here is how to import EPS art
\caption{Schematic of Pt/Au catalytic nanomotor showing electrochemical reactions, generated charge density (red and blue denote high and low charge density, respectively), and approximate electric field lines.  Although a gold-platinum nanomotor is shown here, the results in this work are applicable to other bimetallic combinations that undergo electrochemical reactions~\cite{wang_bipolar_2006}.  As shown here, this nanomotor's autonomous motion would be directed to the right.}
\end{figure}

Figure 1 shows a schematic of a Pt/Au bimetallic nanomotor in an aqueous hydrogen peroxide solution.  Peroxide oxidation at the Pt end generates protons, electrons, and oxygen molecules.  The electrons conduct through the rod to the Au end and complete the reduction reaction by combining with protons, peroxide and oxygen to generate water.  This rod effectively acts like a short-circuited galvanic cell which drives a net migration of protons in the surrounding solution from the anode (Pt) to the cathode (Au). Wang et al. used this basic description along with electrochemical measurements~\cite{wang_bipolar_2006}  to successfully predict the direction of motion of a nanorod composed of any 2 of 6 noble metals.  Although there is a growing consensus that this mechanism is operative in causing the rods' motion, the underlying physical details are not well understood.

Here we provide a model that shows that the asymmetric reactions result in an excess and depletion of protons in the surrounding electrolyte at the anode and cathode ends, respectively.  The proton imbalance results in asymmetric free charge density, as shown in Fig. 1, which generates an electric dipole and field pointing from the anode to the cathode.  The self-generated electric field couples with the charge density to produce an electrical body force that drives fluid from the anode to the cathode.  The fluid motion results in locomotion of the nanowire in the direction of the anode. This physics is similar to electrophoresis, except here the electric field and charge density are induced by the surface reactions.  

\begin{figure*}
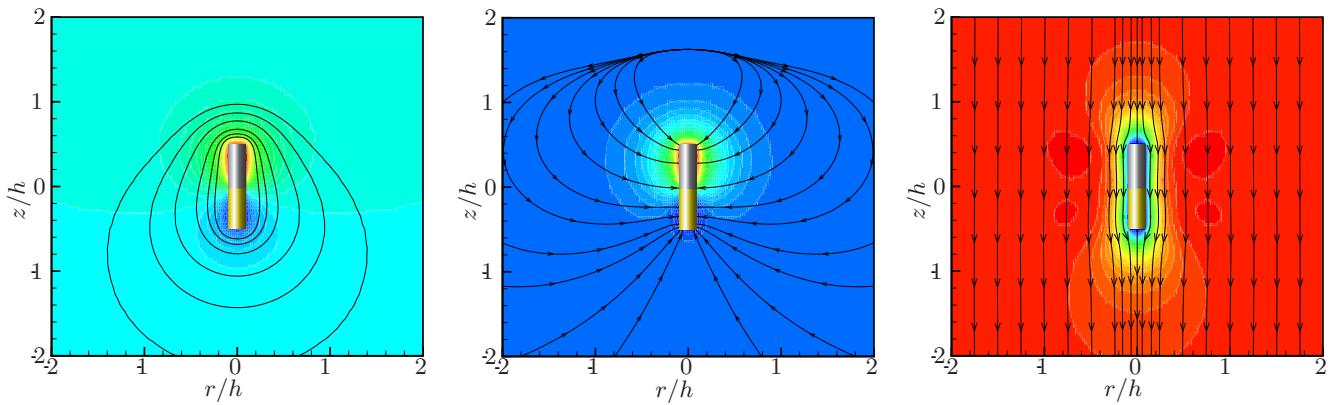

\input{fig2a.tex}
\input{fig2b.tex}
\input{fig2c.tex}
\caption{Simulation-generated plots of (left) normalized proton concentration $c^*$ (color) and electrical potential $\phi^*$(contour lines) and (middle) charge density $\rho_e / \rho_{e_0}$ (color) and electric field (streamlines) and (right)  RICA velocity magnitude (colors) and streamlines (black lines) for the case  $\zeta = -10$ mV and $j / j_d = 0.8$.  The reactions lead to an asymmetry in the proton concentration such that an excess of protons builds up at the anode and protons are depleted at the cathode.  The excess of protons results in positive charge density at the anode and the generation of electric field pointing from the anode to the cathode.  }
\end{figure*}

In our model, we consider a rod immersed in a binary electrolyte with equal concentrations of H$^+$ and OH$^-$ ions.  In the dilute solution limit, the steady ion concentration distributions are given by the advection-diffusion equation, 
\begin{equation} 
 \boldsymbol{u} \cdot \nabla c_i = D_i \nabla^2 c_i + z_i F \nu_i \nabla \cdot (c_i \nabla \phi) 
 \end{equation}
where $\boldsymbol{u}$ is the fluid velocity, $c$ is the molar concentration, $D$ is the diffusivity, $z$ is the valence, $F$ is the Faraday constant, $\nu$ is the mobility, $\phi$ is the electrostatic potential, and the subscript $i$ denotes the species.  The electrostatic potential in turn depends on the local free charge density as described by the Poisson equation,
\begin{equation}    -\epsilon \nabla^2 \phi = \rho_e = F (z_+ c_+ +  z_- c_-)\end{equation}
where $z_+ = 1$, $z_- = -1$, $\rho_e$ is the volumetric charge density, and $\epsilon = \epsilon_r \epsilon_0$ is the permittivity of the liquid which we assume to be constant.   In our model, fluid motion (and thus nanomotor motion) is driven by electrical body forces, which depend on the concentrations of charged species only.  We have estimated the forces due to diffusiophoresis and found that they are several orders of magnitude smaller than those produced by the induced charge mechanism described here.  For this reason we do not consider the oxygen and peroxide concentrations and focus on a simple binary electrolyte.  To close the system of equations we include the steady, incompressible Navier-Stokes equations for a Newtonian fluid,
\begin{eqnarray}
\nabla \cdot &\boldsymbol{u}  =0\\
\rho(\boldsymbol{u} \cdot \nabla \boldsymbol{u} )  = &-\nabla p& + \eta \nabla^2 \boldsymbol{u} - \rho_e \nabla \phi
\end{eqnarray}
Here $\rho$ is the fluid density, $p$ is the pressure, $\eta$ is the dynamic viscosity, and $\rho_e \nabla \phi$  is the electrical body force that results from the coupling of charge density and electric field.  This general framework was developed by Melcher and has been used extensively to describe electrohydrodynamic flows, particularly electrokinetic flows which are driven by a coupling of externally applied fields and charged objects~\cite{hoburg_internal_1976}.  
%In typical electrokinetic models, the electric potential distribution in the solution due to the native surface charge of an object is decoupled from the electric potential gradients that generate electrical body forces~\cite{saville_electrohydrodynamics:taylor-melcher_1997}.   In contrast, in RICA reactions generate charge density that result in electric fields and a net body forces that  drive fluid motion.

The reactions are represented by boundary conditions specifying the molar proton fluxes on the surface of the nanomotor.  On the anode and cathode we prescribe equal and opposite fluxes $j$ and $-j$ normal to the wire surface. Since anions (hydroxide ions) do not participate in the reactions, the normal anion flux is set to zero everywhere on the nanomotor surface.  The values of the proton fluxes specified in the simulations are based on previously published measurements of current density at Pt and Au electrodes in hydrogen peroxide~\cite{laocharoensuk_carbon-nanotube-induced_2008,calvo-marzal_electrochemically-triggered_2009,paxton_catalytically_2006}.  Here we do not directly model the electrochemical reactions that are described by the  Butler-Volmer equation, because we have direct measurements of the current density for our electrocatalysts and fuel.  At the nanomotor surface we apply the no-slip condition for the velocity and specify the local surface potential (relative to the bulk solution) as $\phi=\zeta$.  Far from the rod surface, the electrostatic potential decays to zero and ion concentrations approach their bulk value, i.e.,   $\phi \rightarrow 0$ and $c_i \rightarrow c_\infty$ as the radial distance $r \rightarrow \infty$.

We non-dimensionalize the momentum equation using the following scaling quantities:   $|\boldsymbol{u}| \propto U_{ev}$,  $p \propto \eta U_{ev}/d$, $\rho_e \propto \rho_{e_0}$, and $\nabla \phi \propto E_0$   where $\rho_{e_{0}}$ is a characteristic charge density, $E_0$ is a characteristic electric field,  $d$ is a viscous length scale, and $U_{ev}$ is a characteristic electroviscous velocity.  Applying these scalings to the momentum equation (4), a Reynolds number emerges based on the electroviscous velocity, given by   $\textup{Re} = \rho U_{ev} d / \eta$. Here we use the electroviscous velocity which arises in electrokinetic systems due to the balance of electrical body and viscous forces acting on the fluid, defined as~\cite{hoburg_internal_1976}
\begin{equation}  
U_{ev}  \equiv \frac{\rho_{e_{0}} E_0}{\eta / d^2}. 
\end{equation}
Metallic nanorods support a native surface charge in aqueous solutions \cite{dougherty_zeta_2008}.  The charged surface attracts a screening cloud of counter-ions which develops a region of net charge density in the electrical double layer (EDL) surrounding the rod.  The characteristic length scale of the charge density region is the Debye thickness $\lambda_D$, which scales with the background electrolyte concentration ${c_\infty}^{-1/2}$.  This charge density in the EDL scales with the potential in the EDL based on the Poisson-Boltzmann equation for a symmetric binary electrolyte,
\begin{equation}  
\rho_{e_0}  \propto \frac{2 z^2 F^2 c_\infty \zeta}{RT} \propto \frac{\epsilon\zeta}{\lambda_D^2},
\end{equation}
where we have imposed the Debye-H\"{u}ckel approximation.  In order to also include the effects of the reaction-driven flux, we introduce a characteristic electric field based on the flux and diffusivity of protons,
\begin{equation}  
E_0 \propto \frac{F \lambda_D h}{\epsilon D_+} j,
\end{equation} 
where  $h$ is the length of the nanomotor and also the characteristic length for the electric field.  Combining expressions (5-7), the electroviscous velocity scales as
\begin{equation}  U_{ev} \propto \frac{\zeta F h \lambda_D}{ \eta D_+} j .\end{equation}

Here we scale the viscous  length scale $d$  with the EDL thickness $\lambda_D$  since the region of significant viscous and electrical body forces is limited to the area with charge density.  As will be shown below, the $z$-component of the self-generated electric field is significant along a distance that scales with the nanomotor length $h$. The scaling analysis shows that the nanomotor speed increases linearly with the reaction flux $j$ because the flux generates charge density which produces the internally generated field.   Equation (8) can be recast in the form of the Helmholtz-Smoluchowski equation $U_{ev} = \epsilon \zeta E_0 / \eta$  which describes the electrophoretic velocity for a charged particle in the presence of an external electric field \cite{henry_cataphoresis_1931}, where here $E_0$ is given by equation (7).

We assume the flow is axisymmetric and thus solve the system over a two-dimensional cross-section of the 3-D problem.  The $100 \mu$m $\times 100 \mu$m  simulation domain is discretized into approximately 181,000 triangular mesh elements.  The length and diameter of the simulated nanomotor are set to 2 $\mu$m and 370 nm, respectively.  We solve the system of governing equations (1-4) numerically using the linear system solver  {\sc pardiso}.  Using a two-ion model with a fixed bulk electrolyte concentration, the only free parameters in the system are the nanomotor native surface potential $\zeta$ and the surface flux $j$.  We normalize the flux by characteristic flux based on Nernst's diffusion limited current density given by $j_d = 4 D_+ c_\infty / \lambda_D$~\cite{bazant_current-voltage_2005}.

Figure 2(a) shows the dimensionless proton concentration $c^* = (c_+ - c_\infty)/c_\infty$  (color) and contours of the electric potential normalized by the thermal voltage $\phi^* = z F \phi / RT$ (black lines) for the case where $\zeta = -10$~mV and $j / j_d$ = 0.8.  In the absence of reactions, the EDL proton concentration and electrical potential are both symmetric around the nanorod.  Figure 2(a) shows that when reaction-driven fluxes are introduced, an asymmetry in the proton concentration is established such that an excess of protons builds up at the anode and protons are depleted at the cathode.  The reactions also result in an asymmetric electrical potential profile that bulges at the cathode.

Figure 2(b) shows the normalized charge density and streamlines of electric field for the same case as Fig. 2(a).   The charge density in the diffuse layer of the anode is positive because the negatively charged surface attracts cations near the surface and the surface reactions constantly inject cations. At the cathode the deficiency of  protons due to reactions and the shielding protons due to the negatively charged surface nearly counteract each other resulting in weak negative charge density.   The charge density  generates an electric field, as described mathematically by Poisson's equation.  The electric field couples with the charge density to produce an  electrical body force, $\rho_e \boldsymbol{E}$, which acts on the fluid to drive an electroviscous velocity and propel the nanomotor.

Figure 2(c) shows the RICA velocity magnitude (color) and streamlines (black lines).  These simulations are conducted in the reference frame of the nanomotor.  Fluid flows from the anode to the cathode due to electrical body forces that result from a coupling of positive charge density and electric field tangent to the nanomotor surface.  By Galilean invariance, this is equivalent to the nanomotor swimming with the anode end forward.

\begin{figure}
% This file is generated by the MATLAB m-file laprint.m. It can be included
% into LaTeX documents using the packages graphicx, color and psfrag.
% It is accompanied by a postscript file. A sample LaTeX file is:
%    \documentclass{article}\usepackage{graphicx,color,psfrag}
%    \begin{document}\input{newfig3}\end{document}
% See http://www.mathworks.de/matlabcentral/fileexchange/loadFile.do?objectId=4638
% for recent versions of laprint.m.
%
% created by:           LaPrint version 3.16 (13.9.2004)
% created on:           29-Jan-2010 10:39:15
% eps bounding box:     15 cm x 11.25 cm
% comment:              
%
\begin{psfrags}%
\psfragscanon%
%
% text strings:
\psfrag{s05}[t][t]{\color[rgb]{0,0,0}\setlength{\tabcolsep}{0pt}\begin{tabular}{c}$j/j_d$\end{tabular}}%
\psfrag{s06}[b][b]{\color[rgb]{0,0,0}\setlength{\tabcolsep}{0pt}\begin{tabular}{c}$U_{ev}$ ($\mu$m/s)\end{tabular}}%
\psfrag{s10}[l][l]{\color[rgb]{0,0,0}Experiments}%
\psfrag{s11}[l][l]{\color[rgb]{0,0,0}$\zeta$ = -40 mV}%
\psfrag{s12}[l][l]{\color[rgb]{0,0,0}$\zeta$ = -30 mV}%
\psfrag{s13}[l][l]{\color[rgb]{0,0,0}$\zeta$ = -20 mV}%
\psfrag{s14}[l][l]{\color[rgb]{0,0,0}$\zeta$ = -10 mV}%
\psfrag{s15}[l][l]{\color[rgb]{0,0,0}Experiments}%
%
% xticklabels:
\psfrag{x01}[t][t]{0}%
\psfrag{x02}[t][t]{0.1}%
\psfrag{x03}[t][t]{0.2}%
\psfrag{x04}[t][t]{0.3}%
\psfrag{x05}[t][t]{0.4}%
\psfrag{x06}[t][t]{0.5}%
\psfrag{x07}[t][t]{0.6}%
\psfrag{x08}[t][t]{0.7}%
\psfrag{x09}[t][t]{0.8}%
\psfrag{x10}[t][t]{0.9}%
\psfrag{x11}[t][t]{1}%
\psfrag{x12}[t][t]{0}%
\psfrag{x13}[t][t]{0.5}%
\psfrag{x14}[t][t]{1}%
\psfrag{x15}[t][t]{1.5}%
\psfrag{x16}[t][t]{2}%
%
% yticklabels:
\psfrag{v01}[r][r]{0}%
\psfrag{v02}[r][r]{0.1}%
\psfrag{v03}[r][r]{0.2}%
\psfrag{v04}[r][r]{0.3}%
\psfrag{v05}[r][r]{0.4}%
\psfrag{v06}[r][r]{0.5}%
\psfrag{v07}[r][r]{0.6}%
\psfrag{v08}[r][r]{0.7}%
\psfrag{v09}[r][r]{0.8}%
\psfrag{v10}[r][r]{0.9}%
\psfrag{v11}[r][r]{1}%
\psfrag{v12}[r][r]{0}%
\psfrag{v13}[r][r]{5}%
\psfrag{v14}[r][r]{10}%
\psfrag{v15}[r][r]{15}%
\psfrag{v16}[r][r]{20}%
\psfrag{v17}[r][r]{25}%
\psfrag{v18}[r][r]{30}%
%
% Figure:
\includegraphics[clip,width=3.1in]{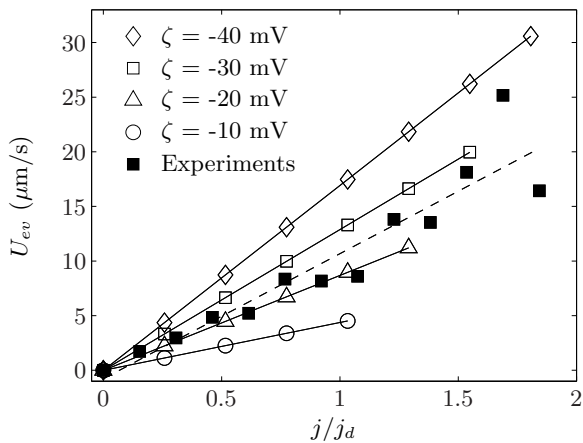}
\end{psfrags}%
%
% End newfig3.tex

\caption{Nanomotor velocity as a function of dimensionless flux  $j / j_d$.  Simulations (open symbols), the scaling analysis (lines), and experiments (closed symbols) show excellent agreement.  Simulations are shown for four values of the zeta potential.   }
\end{figure}

Figure 3 shows the nanomotor velocity as a function of the flux $j / j_d$ obtained from simulations, the scaling analysis, and experiments.  The experiments and simulations show good agreement for a native surface potential of  $-25$~mV, which we obtain from independent measurements of the zeta potential for Au and Pt particles in aqueous solutions ~\cite{dougherty_zeta_2008,du_controlled_2008}.  The nanomotors used in the experiments were grown using electrochemical deposition~\cite{laocharoensuk_carbon-nanotube-induced_2008}. The dimensions of these nanomotors were similar to those simulated here.  We obtained the experimental data in Fig. 3 by measuring the velocity of approximately 200 nanomotors using optical microscopy in varying concentrations of hydrogen peroxide.  The value of $j / j_d$ at each peroxide concentration is estimated from the published dependence of electrocatalytically generated current density on Pt and Au interdigitated microelectrodes~\cite{paxton_catalytically_2006}. We have subtracted out the characteristic Brownian velocity of the nanomotors (measured here to be 4.87 $\mu$m s$^{-1}$) from all experimental data points in order to only consider the axial velocities measured in the experiments.

Additional simulations (not shown) show that the nanomotor velocity scales inversely with $\eta$ and $D$ and directly with the body force as predicted by the scaling analysis.  Viscosity  slows the nanomotor due to Stokes drag and diffusion tends to  reduce the proton concentration gradients generated by the reactions thereby reducing the generated electric fields.  We compute the total propulsive force by numerically integrating the $z$-component of electrical body force, $\rho_e E_z$, over the entire simulation domain and  find that the velocity scales linearly with this body force, in agreement with StokesÕ law.  We calculate a force of 0.17 pN at $j / j_d = 1.0$, $\zeta = -20$ mV  which is in good agreement with our previous experimental measurement of nanomotor propulsive force of 0.16 pN~\cite{burdick_synthetic_2008}.

%One important element is the effect of the background electrolyte concentration on the nanomotor velocity.  It has been reported that the velocity of bimetallic nanomotors typically decreases with increasing electrolyte concentration~\cite{kagan_chemical,paxton_catalytically_2006}.  Our scaling analysis predicts a linear dependence of electroviscous velocity on the Debye length or $c_\infty^{-1/2}$.  We have conducted simulations that indicate this relationship,  although the results have not been reported here.  

We have presented detailed simulations, scaling analysis, and experiments that describe the locomotion of bimetallic nanomotors in hydrogen peroxide solutions due to Reaction Induced Charge Auto-Electrophoresis.  Nanomotor movement is the result of an electroviscous slip velocity that is driven by electrical body forces resulting from charge density and electric fields that are internally generated by electrochemical reactions occurring on the particle surface.  We expect that a detailed understanding of the physics underlying the nanomotors' motion will provide a basis for rational design of next-generation nanomachines capable of operation in diverse conditions and applications.

The authors acknowledge  Kamil Salloum, Marcus Herrmann, and Joseph Wang for stimulating discussions. This work was sponsored by NSF graduate fellowships to JLM and PMW and grant CBET-0853379.

\bibliography{nm}% Produces the bibliography via BibTeX.

\end{document}